\documentclass[amsmath,amssymb,prl,final,superscriptaddress,showpacs,twocolumn]{revtex4-1}
 
\usepackage{bm,hyphenat,xspace}
\usepackage{graphicx,epsfig}

\newcommand {\mcu}{\mathcal{U}}

\newcommand{\cm}{\mathrm{c\!\:\!.m\!\:\!.}}

\newcommand{\He}{{}^3\mathrm{He}}
\newcommand{\Hh}{{}^3\mathrm{H}}
\newcommand{\nH}{n\text{-}{}^3\mathrm{H}}
\newcommand{\pHe}{p\text{-}{}^3\mathrm{He}}
\newcommand{\pH}{p\text{-}{}^3\mathrm{H}}
\newcommand{\nHe}{n\text{-}{}^3\mathrm{He}}
\newcommand{\pd}{p\text{-}d}

\newcommand{\dd}{d\text{-}d}

\begin{document}
 
\title {Calculation of multichannel reactions in the four-nucleon system above breakup threshold}
 
\author{A.~Deltuva} 
\affiliation{Centro de F\'{\i}sica Nuclear, Universidade de Lisboa, 
P-1649-003 Lisboa, Portugal }
\affiliation{Institute of Theoretical Physics and Astronomy, 
Vilnius University, A. Go\v{s}tauto 12, LT-01108 Vilnius, Lithuania}

\author{A.~C.~Fonseca} 
\affiliation{Centro de F\'{\i}sica Nuclear, Universidade de Lisboa, 
P-1649-003 Lisboa, Portugal }

\received{16 June 2014}
\pacs{21.45.-v, 25.10.+s, 21.30.-x, 24.70.+s}

\begin{abstract}
Exact four-body equations of Alt, Grassberger and Sandhas are solved
for neutron-${}^3\mathrm{He}$ and proton-${}^3\mathrm{H}$ scattering
in the energy regime above the four-nucleon breakup threshold.
Cross sections and spin observables for
elastic, transfer, charge-exchange, and breakup reactions are
calculated using realistic nucleon-nucleon 
interaction models, including the one with effective many-nucleon
forces due to explicit $\Delta$-isobar excitation.
The experimental data are described reasonably
well with only few exceptions such as vector analyzing powers.
\end{abstract}

 \maketitle


Collisions and reactions are among
the most important processes  used 
to study  various quantum systems ranging
from ultracold atoms to nuclear and particle physics.
However, reliable information about their properties
can be extracted from experimental data only when an accurate 
theoretical description of the scattering process is available
which is much more complicated to obtain as compared to bound systems.
With the three-particle system already well under control
the next step is the study of collisions and reactions involving four-particles.

Due to its inherent complexity, rich structure of resonances, and
multitude of channels 
the four-nucleon ($4N$) system constitutes a highly challenging but 
also promising theoretical laboratory to
test the nucleon-nucleon ($NN$) force models. But for that to be possible 
one needs to be able to solve numerically, over a broad range of
energy, the corresponding scattering equations in
momentum or coordinate space. Work
on this problem evolved slowly over the years but  took a fast leap
forward in the last ten years.

Four-nucleon scattering results with realistic force models
emerged first through coordinate space calculations but were limited
to single channel $\nH$ and $\pHe$
 \cite{viviani:01a,kievsky:08a,lazauskas:04a} 
and $\pH$ \cite{lazauskas:09a} reactions below inelastic threshold.
In that region the $4N$ system 
exhibits a rich structure of resonances \cite{tilley:92a} in
different partial waves that have been well identified in the
literature and whose understanding in terms of the underlying force
models constitutes a major yet unresolved challenge for theory. More recent
results show that adding a three-nucleon ($3N$)  force 
 does not necessarily improve the agreement with the experimental data
\cite{lazauskas:04a,lazauskas:05a,fisher:06},
unless particular $3N$ force models are used \cite{viviani:13a}.
As in the three-nucleon system, complex scaling
methods are now being used to calculate single channel reactions above
breakup threshold, however, with semirealistic $S$-wave 
interactions  so far \cite{lazauskas:12a}.

Given that the treatment of the Coulomb interaction between protons
became possible in momentum-space calculations by using the method of
screening and renormalization \cite{deltuva:05a},  solutions of the Alt,
Grassberger and Sandhas (AGS) equations \cite{grassberger:67}
for the transition operators have been accomplished at energies below breakup
threshold for a number of realistic $NN$ interactions \cite{deltuva:07a}.
Because asymptotic
boundary conditions are naturally imposed by the way one handles the
two-cluster singularities, one could calculate cross sections and spin
observables for all two-cluster reactions ranging from $\nH$
\cite{deltuva:07a}, $\pHe$ \cite{deltuva:07b}, $\nHe$, $\pH$ and $\dd$
\cite{deltuva:07c} elastic scattering to rearrangement reactions such as
${}^3\mathrm{H}(p,n){}^3\mathrm{He}$,
${}^2\mathrm{H}(d,p){}^3\mathrm{H}$, and
${}^2\mathrm{H}(d,n){}^3\mathrm{He}$ \cite{deltuva:07c} and their
respective time reversal. In this energy range calculations were done
using real-axis integration with subtraction method, spline interpolation, 
 and Pad\'{e}  summation \cite{baker:75a} for matrix elements of
the  transition operators \cite{deltuva:07a}. 
This approach was extended 
to allow for an explicit $\Delta$-isobar excitation yielding
effective $3N$ and $4N$ forces \cite{deltuva:08a}.
Unlike coordinate space methods, adding an irreducible
$3N$ force constitutes a major stumbling
block for momentum-space calculations that has not yet been resolved,
except for bound state calculations \cite{nogga:02a}.

Given the complex analytical structure of the four-body kernel in
momentum space above three- and, especially, four-cluster threshold, 
going beyond breakup
threshold seemed for a while an impossible endeavor. 
 Using complex energy in the form of $Z = E + i\epsilon$, where
$\epsilon$ is a finite quantity  \cite{uzu:03a}, was a mirage that
only worked well when a new integration method with special weights 
\cite{deltuva:12c} was developed taking into account the
presence of the quasisingularities. 
 This enabled fully realistic state-of-the-art
calculations of $\nH$ \cite{deltuva:12c} and $\pHe$ \cite{deltuva:13c}
elastic scattering up to 35 MeV nucleon energy. 

In this work we present the most challenging step in $4N$
 scattering research by addressing $\nHe$ and $\pH$
mixed isospin $T=0$ and 1 reactions above the
breakup threshold where the singularity structure of the
four-nucleon kernel attains its highest complexity
due to a variety of open channels.
This continues the investigations we pioneered years ago below breakup
\cite{deltuva:07c}.
Because all two, three- and four-cluster channels are
open we are confronted with the most complex
nuclear reaction calculation to date that actually resembles a typical nuclear
reaction process where elastic, charge exchange, transfer and breakup
take place simultaneously. 

We  use the symmetrized AGS equations \cite{deltuva:07a} as
appropriate for the four-nucleon system in the isospin formalism.
They are integral equations for the four-particle transition operators
$\mcu_{\beta \alpha}$, i.e.,
\begin{subequations}  \label{eq:AGS}   
\begin{align}  
\mcu_{11}  = {}&  -(G_0 \, t \, G_0)^{-1}  P_{34} - P_{34} U_1 G_0 \,
t \, G_0 \, \mcu_{11}  \nonumber \\  {}& + U_2   G_0 \, t \, G_0 \,
\mcu_{21}, \label{eq:U11}  \\
\label{eq:U21}
\mcu_{21} = {}&  (G_0 \, t \, G_0)^{-1}  (1 - P_{34}) + (1 - P_{34})
U_1 G_0 \, t \, G_0 \, \mcu_{11}.
\end{align}
\end{subequations}
Here, $\alpha=1$ corresponds to the $3+1$ partition (12,3)4 whereas
$\alpha=2$ corresponds to the $2+2$ partition (12)(34); there are no
other distinct  two-cluster partitions in the system of four identical
particles.  The symmetrized 3+1 or 2+2 subsystem transition operators
are obtained from the respective integral equations
\begin{gather} \label{eq:AGSsub}
U_\alpha =  P_\alpha G_0^{-1} + P_\alpha t\, G_0 \, U_\alpha.
\end{gather}
The free resolvent at the complex energy $E+ i\varepsilon$ is given by
\begin{gather}\label{eq:G0}
G_0 = (E+ i\varepsilon - H_0)^{-1},
\end{gather} 
with  $H_0$ being the free Hamiltonian.  The pair (12) transition
matrix
\begin{gather} \label{eq:t}
t = v + v G_0 t
\end{gather} 
is derived from the potential $v$; for the $pp$ pair  $v$ includes
both the nuclear and the  screened Coulomb potential.  Thereby
all transition operators acquire parametric dependence on 
the screening radius $R$ but for simplicity it
is suppressed in our notation. 
The full antisymmetry of the four-nucleon system is ensured by the
permutation operators $P_{ab}$ of particles $a$ and $b$ with $P_1 =
P_{12}\, P_{23} + P_{13}\, P_{23}$ and $P_2 =  P_{13}\, P_{24}$,
together with the special symmetry of basis states as pointed out in
Refs.~\cite{deltuva:07a,deltuva:12c}.

The numerical calculations are performed for complex energies, i.e.,
with finite  $\varepsilon$.  A special
integration method developed in  Ref.~\cite{deltuva:12c} is used  to
treat the quasisingularities of the AGS equations \eqref{eq:AGS}.
The limit  $\varepsilon \to +0$ needed for
the calculation of scattering amplitudes and observables
 is obtained by the extrapolation of finite
$\varepsilon$ results. Beside the point method 
proposed in Ref.~\cite{schlessinger:68} and applied in Refs.
\cite{uzu:03a,deltuva:12c} we use, as an additional accuracy check,
cubic spline extrapolation with nonstandard choice of the boundary
conditions, namely, the one ensuring continuity of the third derivative
\cite{chmielewski:03a}. These two different methods lead to 
indistinguishable results confirming the reliability of the calculations.
We use $ \varepsilon$ ranging from 1 to 2
MeV at the lowest considered energies and from 2 to 4 MeV at the highest
considered energies. About 30 grid points for the discretization of each
momentum variable are used. 

The limit  $\varepsilon \to +0$ is calculated separately for 
each value of the Coulomb screening radius $R$. After that the renormalization 
procedure is performed as described in 
Refs.~\cite{deltuva:05a,deltuva:07b,deltuva:07c} and the results are
checked to be independent of $R$ provided $R$ is large enough.
In the present calculations we found $R$ ranging from 10 to 16 fm
(depending on reaction and energy) to be fully sufficient for convergence.

The results are also fully converged with respect to the partial-wave
expansion. The calculations include  $2N$ partial waves
with  orbital angular momentum $l_x \leq 5$, $3N$
partial waves with spectator orbital angular momentum $l_y \leq 5$ and
total angular momentum $J_y \leq \frac{11}{2}$, $4N$ partial waves with
1+3 and 2+2 orbital angular momentum $l_z \leq 6$, resulting in up to
about 21000 channels for fixed $4N$ total angular momentum and parity.
For some reactions, e.g., $\nHe$ elastic scattering, the partial-wave
convergence is considerably faster, allowing for a reduction in the
employed angular momentum cutoffs.
After the AGS equations are solved, for the calculation of observables
it is sufficient to include only the initial and final 1+3 states with 
$l_z \leq 5$ or even $l_z \leq 4$,
the only exception being the transfer reactions requiring 
$l_z \leq 6$ for both 1+3 and 2+2 channels.
Further  technical details on the solution
of the four-nucleon  AGS equations can be found in
Refs.~\cite{deltuva:07a,deltuva:12c}. 


We study $\nHe$ and $\pH$ scattering  using  realistic high-precision 
two-baryon potentials, namely, 
the purely nucleonic 
inside-nonlocal outside-Yukawa (INOY04) potential  by Doleschall
\cite{doleschall:04a,lazauskas:04a} and the coupled-channel extension 
of the charge-dependent Bonn potential (CD Bonn)  \cite{machleidt:01a}, 
called  CD Bonn + $\Delta$ \cite{deltuva:03c}. The latter
 allows for an excitation of a nucleon to
a $\Delta$ isobar effectively yielding three- and
four-nucleon forces (3NF and 4NF).  The $\He$ ($\Hh$) binding energy
calculated with CD Bonn + $\Delta$ and INOY04
potentials is  7.53 (8.28) and 7.73 (8.49) MeV, respectively; the
experimental value is 7.72 (8.48) MeV.  Therefore most of our predictions
are obtained using INOY04 as it nearly
reproduces the experimental binding energies. The calculations
with CD Bonn + $\Delta$ are done at fewer selected energies. 
Thus, we will compare results obtained with a
pure nucleonic pairwise interaction that by itself alone leads to the
correct $\He$ and $\Hh$ binding, with those obtained with a force
model where effective $3N$ and $4N$ forces are present but does not quite
reproduce the three-nucleon binding.

\begin{figure}[!]
\begin{center}
\includegraphics[scale=0.56]{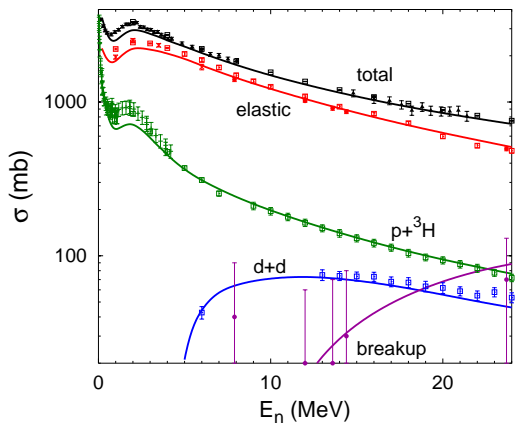}
\end{center}
\caption{\label{fig:nhtot} (Color online) $\nHe$ total and partial
  cross sections as functions  of the neutron beam energy calculated
  using INOY04 potential.  The data  are from 
Refs.~\cite{battat:59,PhysRevC.28.995,drosg:74a,seagrave:60,PhysRev.114.571}.}
\end{figure}

In Fig.~\ref{fig:nhtot} we show the $\nHe$ total and partial 
cross sections for all open channels
as functions of the incoming neutron beam energy $E_n$
ranging from 0 to 24 MeV. Results  obtained using 
 the INOY04 potential are compared with data from 
Refs.~\cite{battat:59,PhysRevC.28.995,drosg:74a,seagrave:60,PhysRev.114.571}.
Below 5 MeV where several resonant $4N$ states are present 
the theory underpredicts  the
data as already pointed out in Ref.~\cite{deltuva:07c}, 
but at higher energies the predictions 
follow the data, not just for the largest
total and elastic cross sections but also for 
the  $\He(n,p)\Hh$ charge exchange  and  the  $\He(n,d)^2\rm{H}$ transfer
reactions. Using unitarity we
calculate also the total breakup cross section 
(three- plus four-cluster) which rises
fast and quickly becomes the dominant inelastic channel; 
the few existing data points have large error bars and, 
except for the first one at $E_n =8$ MeV,
are consistent with our predictions.

\begin{figure}[!]
\begin{center}
\includegraphics[scale=0.6]{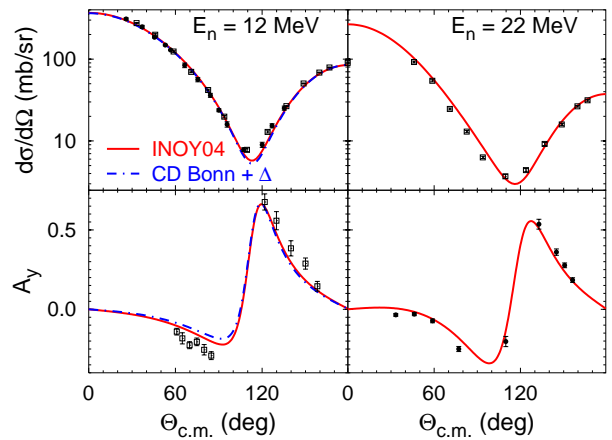}
\end{center}
\caption{\label{fig:nh} (Color online) Differential cross section and
  neutron analyzing power of elastic $\nHe$ scattering at 12 and 22
  MeV neutron  energy.  Results obtained with potentials INOY04
  (solid curves) and CD Bonn + $\Delta$ (dashed-dotted curves) are
  compared with data  from 
Refs.~\cite{haesner:exfor,drosg:74a,lisowski:76a,klages:85}.}
\end{figure}

\begin{figure}[!]
\begin{center}
\includegraphics[scale=0.58]{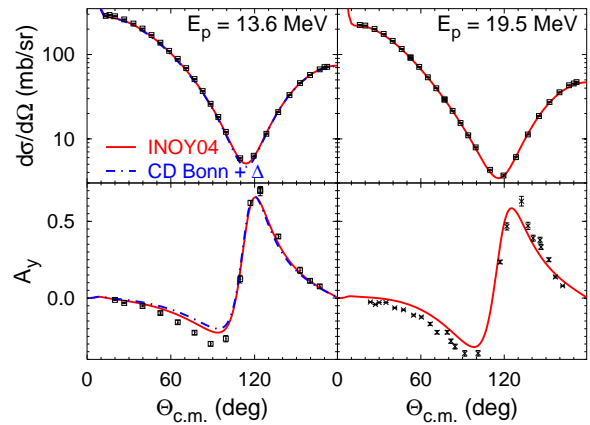}
\end{center}
\caption{\label{fig:pt} (Color online) Differential cross section and
  proton analyzing power of elastic $\pH$ scattering at 13.6 and 19.5
  MeV proton energy.  Curves as in Fig.~\ref{fig:nh}.  The data
  are from Refs.~\cite{PhysRevC.4.52,pt19-57}.}
\end{figure}

In Figs.~\ref{fig:nh} and \ref{fig:pt} we show the
angular dependence of
the differential cross sections $d\sigma/d\Omega$
and nucleon
analyzing powers $A_{y}$ for $\nHe$ and $\pH$ elastic scattering
at few selected energies above the four-cluster breakup threshold.
Differential cross sections peak at forward direction
and show a deep minimum around $\Theta_\cm = 115^\circ$.
Unlike at low energies, analyzing powers develop a shallow
minimum around $90^\circ$   before they rise to a maximum around $120^\circ$.
All these special features in the observables slowly move to larger
angles with increasing energy.
 Much like in  $\pd$ and $\pHe$ elastic scattering we observe that
a reasonable description of the data  
\cite{haesner:exfor,drosg:74a,lisowski:76a,klages:85,PhysRevC.4.52,pt19-57}
emerges as the energy rises;
there remain only small or at most moderate discrepancies,
the most notable one being around the $A_{ y}$ minimum.
At $E_n = 12$ MeV  and
 $E_p = 13.6$ MeV  we show also the  results obtained 
with the CD Bonn + $\Delta$ interaction where effective $3N$ and $4N$ 
forces are included. The difference between the INOY04 and CD Bonn + $\Delta$
only shows up around the
minima of $d\sigma/d\Omega$ and  $A_{ y}$ 
 indicating that the elastic scattering results are not
very sensitive to the force model.

\begin{figure}[!]
\begin{center}
\includegraphics[scale=0.58]{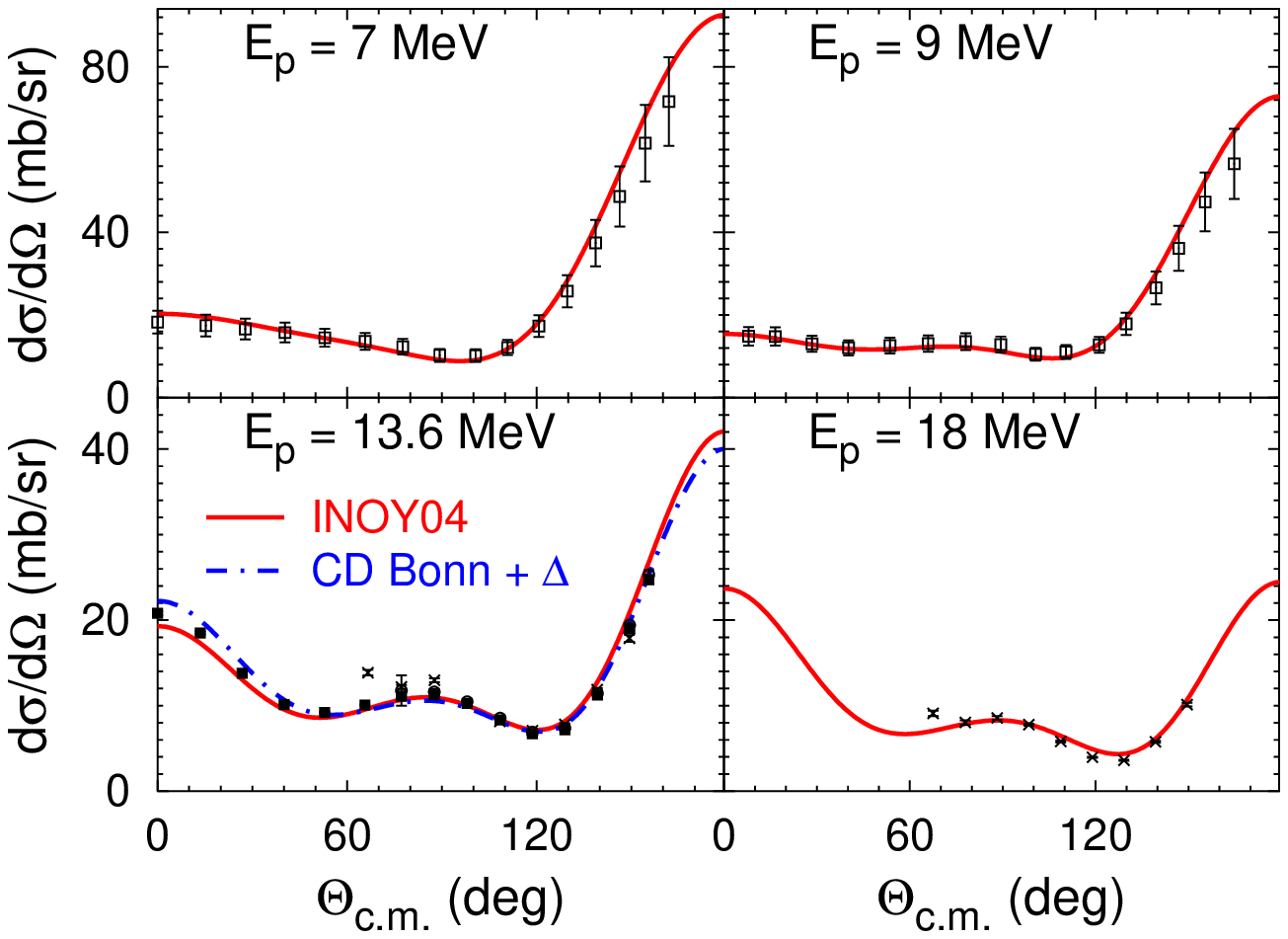}
\end{center}
\caption{\label{fig:pt-nhs} (Color online) Differential cross section
  of  $\Hh(p,n)\He$ reaction at 7, 9, 13.6, and 18 MeV proton
  energy.  Curves as in Fig.~\ref{fig:nh}.  The data  are from
  Refs.~\cite{wilson:61,drosg:78,PhysRevC.16.15,allas:74a}. }
\end{figure}

\begin{figure}[!]
\begin{center}
\includegraphics[scale=0.58]{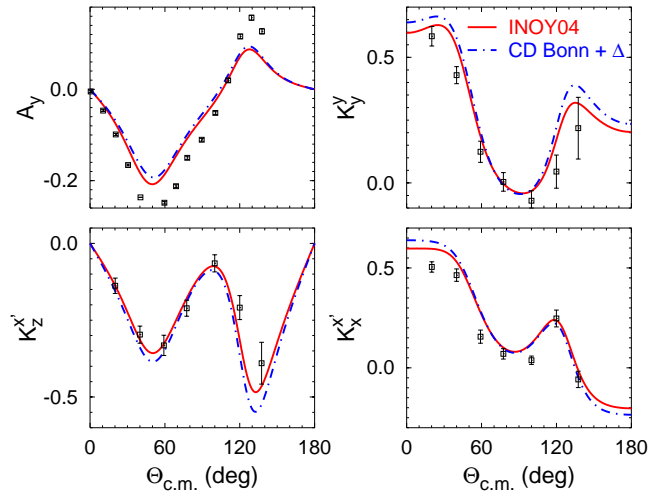}
\end{center}
\caption{\label{fig:pt-nhak} (Color online) Proton analyzing power and
  proton to neutron polarization transfer  coefficients of
  $\Hh(p,n)\He$ reaction at 13.6 MeV proton energy.  Curves as in
  Fig.~\ref{fig:nh}.  
The data are from Refs.~\cite{jarmer:74,PhysRevC.5.1826}.}
\end{figure}

Next we consider 
the charge exchange reaction $\Hh(p,n)\He$ at proton beam
energy $E_p$ ranging from 7 to 18 MeV. In contrast to elastic scattering,
the differential cross section for the charge exchange reaction
exhibits a strong energy dependence as demonstrated in Fig.~\ref{fig:pt-nhs}.
In particular,  a simple shape of $d\sigma/d\Omega$ at $E_p= 7$ MeV
with a single minimum around $\Theta_\cm = 95^\circ$  evolves   
into a more complex one with a local maximum around
$80^\circ$ -  $90^\circ$ and two minima near $50^\circ$ and  $120^\circ$.
 Furthermore, the backward peak decreases rapidly with increasing
energy while the evolution of the forward peak is non-monotonic:
first it slowly decreases but then starts to increase for $E_p > 10$ MeV.
Given such a complicated behavior of the 
differential cross section for the charge-exchange reaction,
the observed good agreement between our theoretical predictions
and experimental data from
 Refs.~\cite{wilson:61,drosg:78,PhysRevC.16.15,allas:74a}
is very impressive. In addition to INOY04 predictions,
at $E_p = 13.6$ MeV we show the results obtained using the CD Bonn + $\Delta$
 interaction model. We find differences of about 10\% and 5\% at 
forward and backward angles, respectively.
Differences between INOY04 and CD Bonn + $\Delta$ 
become even more pronounced in the spin observables, reaching
up to 10 - 15\%.
In Fig.~\ref{fig:pt-nhak}  we show proton analyzing
power $A_{ y}$ and proton to neutron polarization transfer coefficients 
$K_y^y$, $K_z^{x'}$, and  $K_x^{x'}$ for the
charge exchange reaction at $E_p = 13.6$ MeV. The shape
of the observables with several extrema
is reproduced by both interaction models,
but INOY04 yields a better quantitative description of the data
\cite{jarmer:74,PhysRevC.5.1826},
especially for the polarization transfer coefficients.
The extrema of  $A_{ y}$ represent the worst case 
with a significant discrepancy between theory and experiment.
Otherwise, the overall description of the data is impressive
given its complexity and the absence
of any adjustable parameter in the calculations or the interactions 
after they have been fit to the $NN$ data.

\begin{figure}[!]
\begin{center}
\includegraphics[scale=0.6]{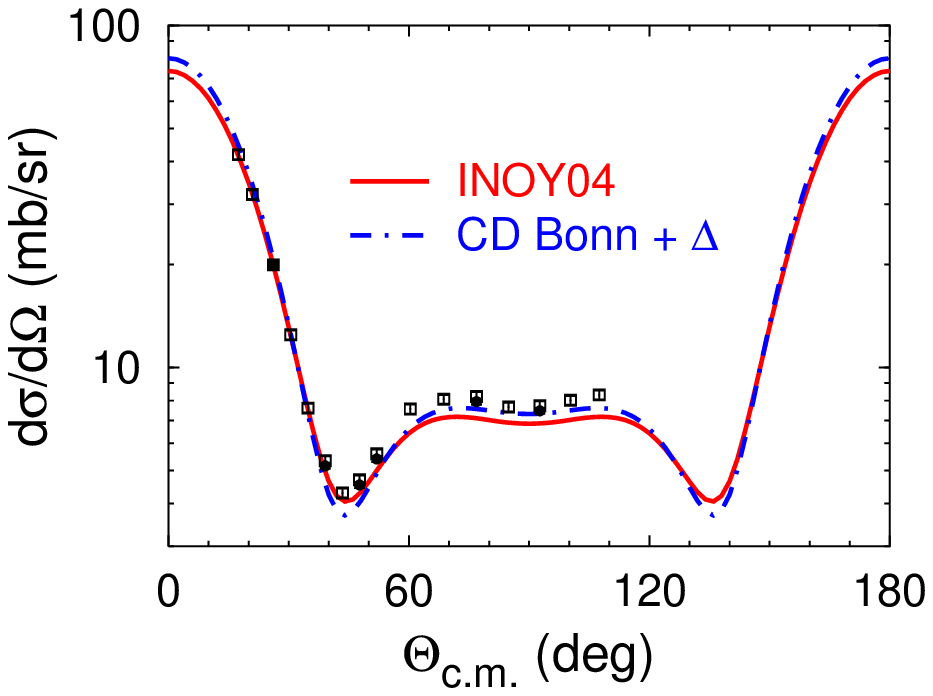}
\end{center}
\caption{\label{fig:pt-dd} (Color online) Differential cross section
  of  $\Hh(p,d){}^2\mathrm{H}$ reaction at 13.6 MeV proton lab energy.
  Curves as in Fig.~\ref{fig:nh}.  The data  are from
  Refs.~\cite{PhysRevC.4.52,PhysRevC.10.54}.}
\end{figure}

Finally in Fig.~\ref{fig:pt-dd} we show the differential cross section for the
$\Hh(p,d)^2\rm{H}$ transfer reaction at 
$E_p = 13.6$ MeV. The shape of the observable having several local extrema
 and a forward peak  
is well reproduced by calculations except at larger angles
 $\Theta_\cm > 45^\circ$ where, after the minimum, the data 
\cite{PhysRevC.4.52,PhysRevC.10.54} are slightly underpredicted. 
The sensitivity to the force model reaches 10\% but this time 
the CD Bonn + $\Delta$ interaction model provides better 
description of the data around  $\Theta_\cm = 90^\circ$.
This sensitivity to some extent may be caused by the correlation
with $\Hh$ binding energy, as found in Ref.~\cite{deltuva:10a}
for ${}^2\rm{H}(d,p)\Hh$ fusion
below three-cluster threshold.

In summary,
the most advanced state-of-the-art calculations involving
all multichannel reactions in the four-nucleon system have been performed
above breakup threshold using exact AGS equations. The 
results were obtained for  mixed isospin ($T=0$ and 1)
reactions initiated by the
scattering of a neutron on a $\He$ target or a proton on a $\Hh$ target.
Realistic nuclear force models INOY04 and CD Bonn + $\Delta$, 
the latter yielding also
effective many-nucleon forces due to $\Delta$-isobar excitation,
  were used together with the proton-proton Coulomb interaction
included via the screening and renormalization method.
Calculations are fully converged with respect to numerical integration,
angular momentum decomposition, and Coulomb screening.

The results are in good agreement with the experimental data 
 except for few observables in specific angular regions.
Even in such cases the overall reproduction of 
the complex structure of the observables having several local extrema
is impressive as demonstrated, e.g., for differential cross section
and polarization transfer coefficients in  the $\Hh(p,n)\He$ reaction. 
The only sizable discrepancy is found for proton analyzing power
in the charge exchange reaction.
Calculations with different potentials show 
that inelastic reactions and spin observables 
 are more sensitive to interaction models.

Further calculations, including elastic, transfer, and breakup reactions
in deuteron-deuteron scattering,  are under way to gain deeper insight 
into the four-nucleon system and the chosen nuclear force models.
This work also opens the door to study the validity of approximate methods 
in reaction theory to treat direct nuclear reactions.
Furthermore, the present developments are also of great importance
to cold-atom physics where the momentum-space AGS equations
were proven to be a powerful tool to study universal
phenomena \cite{deltuva:10c,deltuva:11a}. For example,
implementing the ideas of this
work in the calculations of Ref.~\cite{deltuva:12a} will enable 
the description of four-atom recombination 
at finite temperature.


\begin{thebibliography}{10}

\bibitem{viviani:01a}
M. Viviani {\it et~al.}, Phys. Rev. Lett. {\bf 86},  3739  (2001).

\bibitem{kievsky:08a}
A. Kievsky {\it et~al.}, J. Phys. G {\bf 35},  063101  (2008).

\bibitem{lazauskas:04a}
R. Lazauskas and J. Carbonell, Phys. Rev. C {\bf 70},  044002  (2004).

\bibitem{lazauskas:09a}
R. Lazauskas, Phys. Rev. C {\bf 79},  054007  (2009).

\bibitem{tilley:92a}
D.~R. Tilley, H. Weller, and G.~M. Hale, Nucl. Phys. {\bf A541},  1  (1992).

\bibitem{lazauskas:05a}
R. Lazauskas {\it et~al.}, Phys. Rev. C {\bf 71},  034004  (2005).

\bibitem{fisher:06}
B.~M. Fisher {\it et~al.}, Phys. Rev. C {\bf 74},  034001  (2006).

\bibitem{viviani:13a}
M. Viviani, L. Girlanda, A. Kievsky, and L.~E. Marcucci, Phys. Rev. Lett. {\bf
  111},  172302  (2013).

\bibitem{lazauskas:12a}
R. Lazauskas, Phys. Rev. C {\bf 86},  044002  (2012).

\bibitem{deltuva:05a}
A. Deltuva, A.~C. Fonseca, and P.~U. Sauer, Phys.~Rev.~C {\bf 71},  054005
  (2005).

\bibitem{grassberger:67}
P. Grassberger and W. Sandhas, Nucl. Phys. {\bf B2},  181  (1967); E. O. Alt,
  P. Grassberger, and W. Sandhas, JINR report No. E4-6688 (1972).

\bibitem{deltuva:07a}
A. Deltuva and A.~C. Fonseca, Phys.~Rev.~C {\bf 75},  014005  (2007).

\bibitem{deltuva:07b}
A. Deltuva and A.~C. Fonseca, Phys.~Rev.~Lett. {\bf 98},  162502  (2007).

\bibitem{deltuva:07c}
A. Deltuva and A.~C. Fonseca, Phys.~Rev.~C {\bf 76},  021001(R)  (2007).

\bibitem{baker:75a}
G.~A. Baker, {\em Essentials of Pad\'e Approximants} (Academic Press, New York,
  1975).

\bibitem{deltuva:08a}
A. Deltuva, A.~C. Fonseca, and P.~U. Sauer, Phys.~Lett.~B {\bf 660},  471
  (2008).

\bibitem{nogga:02a}
A. Nogga, H. Kamada, W. Gl\"ockle, and B.~R. Barrett, Phys. Rev. C {\bf 65},
  054003  (2002).

\bibitem{uzu:03a}
E. Uzu, H. Kamada, and Y. Koike, Phys. Rev. C {\bf 68},  061001(R)  (2003).

\bibitem{deltuva:12c}
A. Deltuva and A.~C. Fonseca, Phys.~Rev.~C {\bf 86},  011001(R)  (2012).

\bibitem{deltuva:13c}
A. Deltuva and A.~C. Fonseca, Phys.~Rev.~C {\bf 87},  054002  (2013).

\bibitem{schlessinger:68}
L. Schlessinger, Phys. Rev. {\bf 167},  1411  (1968).

\bibitem{chmielewski:03a}
K. Chmielewski {\it et~al.}, Phys.~Rev.~C {\bf 67},  014002  (2003).

\bibitem{doleschall:04a}
P. Doleschall, Phys.~Rev.~C {\bf 69},  054001  (2004).

\bibitem{machleidt:01a}
R. Machleidt, Phys.~Rev.~C {\bf 63},  024001  (2001).

\bibitem{deltuva:03c}
A. Deltuva, R. Machleidt, and P.~U. Sauer, Phys.~Rev.~C {\bf 68},  024005
  (2003).

\bibitem{battat:59}
M.~E. Battat~{\it et al.}, Nucl. Phys. {\bf 12},  291  (1959).

\bibitem{PhysRevC.28.995}
B. Haesner {\it et~al.}, Phys. Rev. C {\bf 28},  995  (1983).

\bibitem{drosg:74a}
M. Drosg {\it et~al.}, Phys. Rev. C {\bf 9},  179  (1974).

\bibitem{seagrave:60}
J.~D. Seagrave, L. Cranberg, and J.~E. Simmons, Phys. Rev. {\bf 119},  1981
  (1960).

\bibitem{PhysRev.114.571}
J.~H. Gibbons and R.~L. Macklin, Phys. Rev. {\bf 114},  571  (1959).

\bibitem{haesner:exfor}
B. Haesner {\it et~al.},  in {\em EXFOR database} (NNDC, Brookhaven, 1982).

\bibitem{lisowski:76a}
P. Lisowski {\it et~al.}, Nucl. Phys. A {\bf 259},  61   (1976).

\bibitem{klages:85}
H.~O. Klages {\it et~al.}, Nucl. Phys. {\bf A443},  237  (1985).

\bibitem{PhysRevC.4.52}
J.~L. Detch, R.~L. Hutson, N. Jarmie, and J.~H. Jett, Phys. Rev. C {\bf 4},  52
   (1971).

\bibitem{pt19-57}
R. Darves-Blanc {\it et~al.}, Lett. Nuovo Cimento {\bf 4},  16  (1972).

\bibitem{wilson:61}
W.~E. Wilson, R.~L. Walter, and D.~B. Fossan, Nucl. Phys. {\bf 27},  421
  (1961).

\bibitem{drosg:78}
M. Drosg, Nucl. Sci. Eng. {\bf 67},  190  (1978).

\bibitem{PhysRevC.16.15}
N. Jarmie and J.~H. Jett, Phys. Rev. C {\bf 16},  15  (1977).

\bibitem{allas:74a}
R.~G. Allas {\it et~al.}, Phys. Rev. C {\bf 9},  787  (1974).

\bibitem{jarmer:74}
J.~J. Jarmer {\it et~al.}, Phys. Rev. C {\bf 9},  1292  (1974).

\bibitem{PhysRevC.5.1826}
R.~C. Haight, J.~E. Simmons, and T.~R. Donoghue, Phys. Rev. C {\bf 5},  1826
  (1972).

\bibitem{PhysRevC.10.54}
N. Jarmie and J.~H. Jett, Phys. Rev. C {\bf 10},  54  (1974).

\bibitem{deltuva:10a}
A. Deltuva and A.~C. Fonseca, Phys.~Rev.~C {\bf 81},  054002  (2010).

\bibitem{deltuva:10c}
A. Deltuva, Phys.~Rev.~A {\bf 82},  040701(R)  (2010).

\bibitem{deltuva:11a}
A. Deltuva, EPL {\bf 95},  43002  (2011).

\bibitem{deltuva:12a}
A. Deltuva, Phys.~Rev.~A {\bf 85},  012708  (2012).

\end{thebibliography}

\end{document}